\documentclass[twocolumn]{aastex631}

\usepackage{booktabs}

\usepackage{amsmath, balance}
\usepackage{colortbl, multirow}

\begin{document}

\title{\textsc{DRAFTS}: A Deep Learning-Based Radio Fast Transient Search Pipeline}

\correspondingauthor{Yong-Kun Zhang, Di Li, Yi Feng}
\email{ykzhang@nao.cas.cn, dili@mail.tsinghua.edu.cn, yifeng@zhejianglab.com}

\author[0000-0002-8744-3546]{Yong-Kun Zhang}
\affiliation{National Astronomical Observatories, Chinese Academy of Sciences, Beijing 100101, China}

\author[0000-0003-3010-7661]{Di Li}
\affiliation{Department of Astronomy, Tsinghua University, Beijing 100084, China}
\affiliation{National Astronomical Observatories, Chinese Academy of Sciences, Beijing 100101, China}
\affiliation{Research Center for Astronomical Computing, Zhejiang Laboratory, Hangzhou, 311100, China}
\affiliation{New Cornerstone Science Laboratory, Shenzhen, 518054, China}

\author[0000-0002-0475-7479]{Yi Feng}
\affiliation{Research Center for Astronomical Computing, Zhejiang Laboratory, Hangzhou, 311100, China}

\author[0000-0002-9390-9672]{Chao-Wei Tsai}
\affiliation{National Astronomical Observatories, Chinese Academy of Sciences, Beijing 100101, China}
\affiliation{Institute for Frontiers in Astronomy and Astrophysics, Beijing Normal University,  Beijing 102206, China}
\affiliation{University of Chinese Academy of Sciences, Beijing 100049, China}

\author[0000-0002-3386-7159]{Pei Wang}
\affiliation{National Astronomical Observatories, Chinese Academy of Sciences, Beijing 100101, China}
\affiliation{Institute for Frontiers in Astronomy and Astrophysics, Beijing Normal University, Beijing, 102206, China}

\author[0000-0001-6651-7799]{Chen-Hui Niu}
\affiliation{Institute of Astrophysics, Central China Normal University, Wuhan, 430079, China}

\author{Hua-Xi Chen} %
\affiliation{Research Center for Astronomical Computing, Zhejiang Laboratory, Hangzhou, 311100, China}

\author{Yu-Hao Zhu}
\affiliation{National Astronomical Observatories, Chinese Academy of Sciences, Beijing 100101, China}
\affiliation{University of Chinese Academy of Sciences, Beijing 100049, China}

\begin{abstract}

The detection of fast radio bursts (FRBs) in radio astronomy is a complex task due to the challenges posed by radio frequency interference (RFI) and signal dispersion in the interstellar medium. Traditional search algorithms are often inefficient, time-consuming, and generate a high number of false positives. In this paper, we present \textsc{DRAFTS}, a deep learning-based radio fast transient search pipeline. \textsc{DRAFTS} integrates object detection and binary classification techniques to accurately identify FRBs in radio data. We developed a large, real-world dataset of FRBs for training deep learning models. The search test on FAST real observation data demonstrates that \textsc{DRAFTS} performs exceptionally in terms of accuracy, completeness, and search speed. In the re-search of FRB 20190520B observation data, \textsc{DRAFTS} detected more than three times the number of bursts compared to \textsc{Heimdall}, highlighting the potential for future FRB detection and analysis.
\end{abstract}

\keywords{Fast radio bursts --- Deep learning --- Computational methods}

\section{Introduction} \label{sec:intro}

Fast radio bursts (FRBs) have emerged as a new focus of research in radio astronomy, characterized by extremely brief instances of radio pulses \citep{2007Sci...318..777L}. FRBs are of great significance for probing the distribution and evolution of cosmic matter, as well as for the study of fundamental physics \citep{2023RvMP...95c5005Z}. To date, approximately 800 FRB sources have been reported \citep{2023Univ....9..330X}, the vast majority of which have been observed only once, while a few sources exhibit unusual activity \citep{2021Natur.598..267L, 2022RAA....22l4002Z, 2023ApJ...955..142Z}. However, their underlying physical mechanisms remain unclear. With increasing global participation and the introduction of new telescopes, it is expected that tens of thousands of FRBs will be discovered in the near future. This creates the challenge of identifying these short-duration transient luminous events within the vast amounts of radio data. Therefore, the development of an efficient algorithm for real-time FRB detection is of paramount importance.

\begin{figure*}
    \centering
    \includegraphics[width=\textwidth]{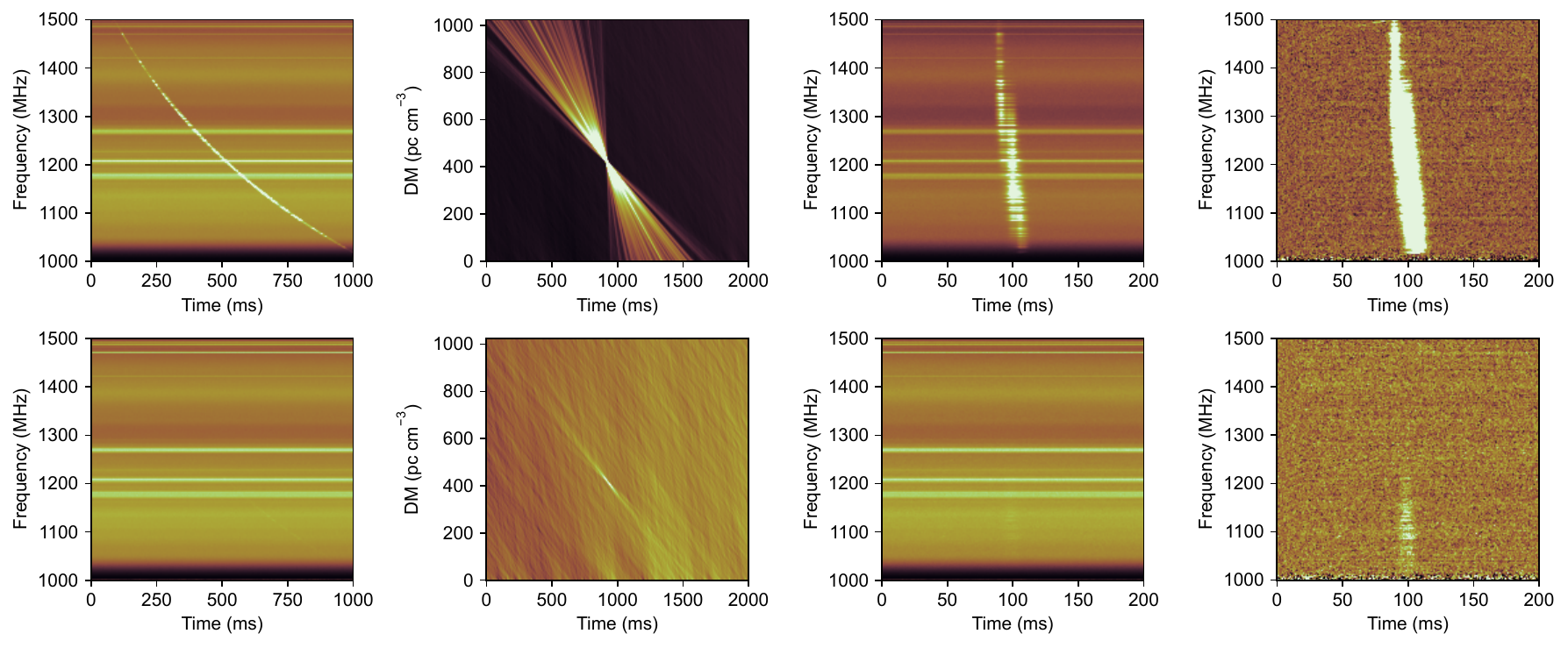}
    \caption{{\bf Radio data of two FRB bursts.} The top and bottom rows show signals of two bursts, one strong and one weak, respectively. From left to right, the first column displays the original data from the radio telescope, with time on the horizontal axis and frequency on the vertical axis. The second column presents the time-DM signals obtained after de-dispersing the original data using a series of DM values. The third column shows the time-frequency data after de-dispersion with the optimal DM value. The fourth column depicts the time-frequency data after RFI mitigation from the data in the third column.}
    \label{fig:radiodata}
\end{figure*}

The search for FRBs can be considered a task of extracting signals with certain characteristics from noise and interference. Fig.~\ref{fig:radiodata} displays two FRB bursts data collected by a radio telescope, with intense radio frequency interference (RFI) present in both data segments, such as the persistent signals near $1200-1300\,\rm MHz$. In addition to the interferences demonstrated in these two data segments, there are more complex types of RFI that actually exist. RFI can be categorized based on variability and frequency range into time-variable RFI/frequency-variable RFI, narrowband interference, and broadband interference. The sources of RFI are diverse, including cell phones, power lines, artificial satellites, lightning, etc., and represent a common problem for all radio telescopes worldwide. Since the interference signals are closer to the telescope than astronomy signals, their intensity is usually orders of magnitude higher than that of astronomy signals, which can obscure the latter.

The characteristic that distinguishes FRB signals from RFI is dispersion. Dispersion occurs as radio signals propagate through the interstellar medium where the speed of signals varies across frequencies, resulting in signals that typically last only a few milliseconds being stretched by a factor of thousands. Consequently, the energy, which would have been concentrated, is dispersed over several seconds of data, causing the signal to become submerged in noise and difficult to detect. The relationship between time delay and frequency is quadratic, forming a parabola, and can be calculated by the following equation:

\begin{equation}
    t_2 - t_1 \approx 4.15 \times \left[\left(\frac{1}{\nu_2}\right)^2 - \left(\frac{1}{\nu_1}\right)^2\right] \times {\rm DM}
\end{equation}
where ${\rm DM}$ represents the parameter that quantifies the amount of plasma along the propagation path of the signal. Hence, radio signals emanating from different locations in the universe will have different ${\rm DM}$ values, leading to varying time delays. We can identify the presence of strong signals from Fig.~\ref{fig:radiodata}, where the time delay of the signal at different frequencies takes on a parabolic form. Bursts originating from different locations in the universe are manifested as parabolas of different curvatures in such data. However, weak bursts are submerged and indistinguishable within the data, and such weak signals are much much more than strong bursts.

\begin{figure*}[!htbp]
    \centering
    \includegraphics[width=\textwidth]{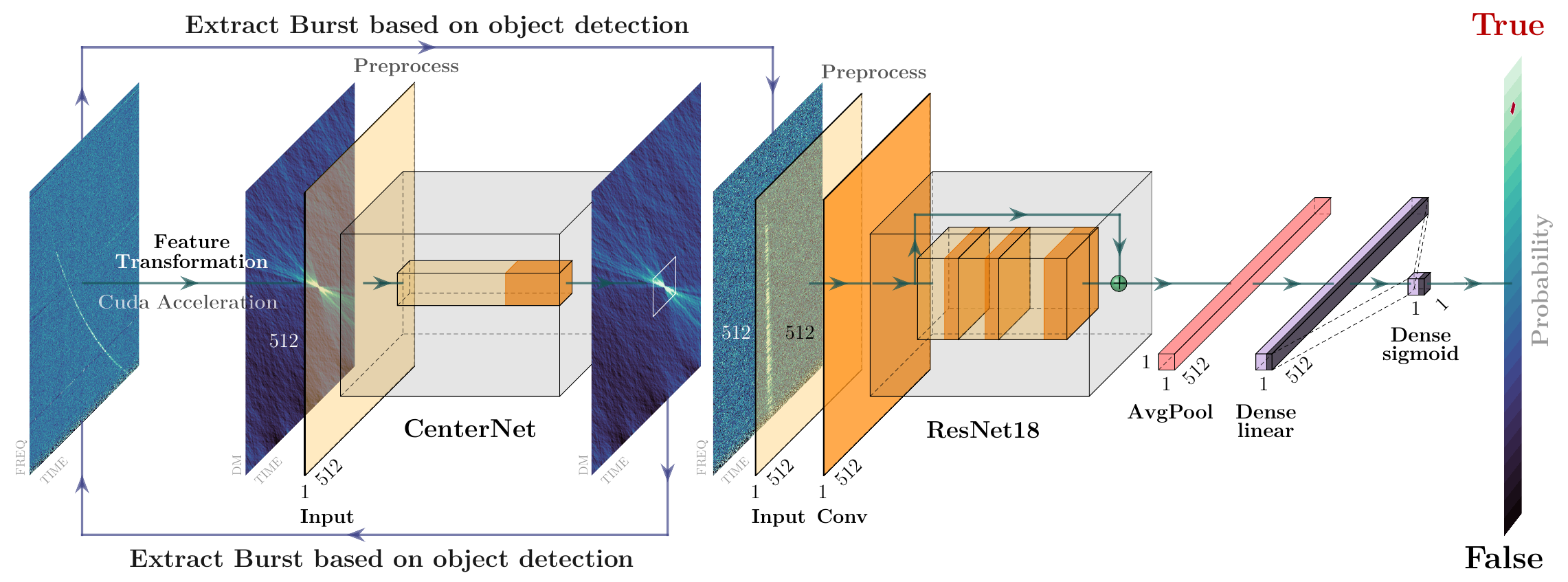}
    \caption{{\bf The workflow of \textsc{DRAFTS}.} }
    \label{fig:search}
\end{figure*}

Therefore, the challenge is to identify faint parabolic-shaped signals from radio data amidst RFI. In this paper, we present \textsc{DRAFTS} \footnote{The code for \textsc{DRAFTS} is available on GitHub at \url{https://github.com/SukiYume/DRAFTS}. \\ The training datasets are available on Hugging Face at \url{https://huggingface.co/datasets/TorchLight/DRAFTS}. \\ The trained models are available on Hugging Face at \url{https://huggingface.co/TorchLight/DRAFTS}.}, a deep learning-based radio fast transient search pipeline, designed for real-time detection of FRBs. \textsc{DRAFTS} is a multi-stage pipeline that integrates object detection and binary classification to efficiently identify FRBs, as illustrated in Fig.~\ref{fig:search}. Trained on an extensive dataset containing both real FRBs and non-FRB signals, \textsc{DRAFTS} demonstrates high performance in terms of detection speed, accuracy, and completeness.

The pipeline's ability to quickly detect FRBs is crucial for advancing scientific discovery, which allows for timely follow-up observations, enhancing our ability to study the dynamic properties of FRBs and their environments \citep{2022A&ARv..30....2P}. Moreover, \textsc{DRAFTS}' high completeness ensures a more thorough detection of faint and rare FRBs, which are critical for understanding the statistical properties of these phenomena \citep{2024SciBu..69.1020Z}. A more complete dataset enables robust statistical analyses, helping to uncover the origins of FRBs and explore their connection to other cosmic processes \citep{2023RvMP...95c5005Z}. This comprehensive approach will support future investigations into the distribution, evolution, and physical mechanisms behind FRBs, ultimately contributing to our broader understanding of the universe.

\section{Related Works}

\subsection{Traditional Methods}

In traditional search algorithms, the process generally includes the following steps \citep{2003ApJ...596.1142C}. Based on a threshold to eliminate RFI, a series of DM grids are set, and the data is de-dispersed accordingly. After that, integration along the frequency direction yields a time series for that DM value. Different widths of boxcar filters are used to match the time series, and the signal-to-noise ratio (S/N) is calculated. Finally, signals that exceed the S/N threshold are selected as candidate signals. Many current tools are based on this process, such as \textsc{Presto} \citep{2001PhDT.......123R} and \textsc{Heimdall} \citep{2012MNRAS.422..379B}. This also constitutes the mainstream method for searching for FRBs.

This method is very intuitive and aligns well with our empirical approach to data processing. However, it strongly depends on both the algorithm used for interference mitigation and the choice of parameters. During data processing, interference that cannot be eliminated might remain, and new artificial data interferences can be introduced as a result of the interference mitigation process. Additionally, the computational complexity is very high, with a lot of redundant calculations, and the same signal can be detected on different DM values, leading to inefficient operation. This ultimately results in incomplete outcomes and the generation of numerous false signals, requiring manual intervention in the data processing procedure and selection of real signals from amongst many false positives. Therefore, this methodology cannot meet the demands of the increasing data volumes expected in the future.

\subsection{Deep Learning Methods}

With the progress of machine learning, especially deep learning in recent years, more researchers are utilizing deep learning methods to address the issues inherent to traditional approaches. In radio astronomy, deep learning has been widely applied to problems such as the classification of radio galaxies \citep{2023MNRAS.522..292B}, the elimination of radio frequency interference \citep{2017A&C....18...35A}, and the reconstruction of radio images \citep{2023arXiv230903291A}. A problem like ours, which involves searching for a specific pattern in two-dimensional data, can be addressed as a computer vision problem.

Some researchers, acknowledging the large number of false signals generated by traditional search methods, have attempted to apply deep learning techniques for the binary classification of detected candidate signals as genuine or spurious \citep{2018AJ....156..256C, 2020MNRAS.497.1661A}. These methods have reduced the manual workload required for signal verification to some extent and have improved the efficiency of the search process. However, they do not address other issues associated with traditional search algorithms, particularly the problem of search incompleteness.

Furthermore, there are also attempts to use deep learning models to directly detect ``parabolic" signals in raw data \citep{2018ApJ...866..149Z, 2022RAA....22j5007L}. However, there are two main problems with this approach. First, weak signals, which are stretched over time, can be inconspicuous in images, making this search method prone to missing such faint signals. The second issue is the variability in the curvature of the parabolic trails in the imagery due to the different DMs associated with FRBs emanating from various locations. This variability makes it challenging to fix the length of input data in blind search, as incomplete coverage of a burst event in the input data could lead the model to overlook the signal. Additionally, even when a signal is detected, the method identifies only its arrival time in the dataset and does not provide dispersion values, calling for further processing to fully characterize the burst.

\subsection{Object Detection}

Object Detection is one of the most critical branches in the field of computer vision and has a wide range of applications in daily life, such as video surveillance and autonomous driving. It aims to understand the visual content within digital images or videos. Object detection not only identifies the categories of objects in the image but also locates their positions within it. In recent years, along with the rapid development of deep learning networks, the performance of object detectors has been significantly improved \citep{jiao2019survey}.

Within the realm of deep learning object detection methods, there are generally two types: anchor-based and anchor-free methods. Anchor-based methods, such as the RCNN series \citep{girshick2014rich, girshickICCV15fastrcnn, renNIPS15fasterrcnn, he2017mask} and YOLO series \citep{redmon2016you, redmon2018yolov3, bochkovskiy2020yolov4, ge2021yolox, li2022yolov6, wang2023yolov7, wang2024yolov9}, operate by predefining a series of fixed boxes (known as anchors) in the image and then predicting the position and categories of objects based on these anchors. Although this method performs well, it has certain limitations. For example, detection performance is highly sensitive to the size, quantity, and aspect ratio of anchors; fixed-size anchors lead to lower detection performance for small-scale objects. Moreover, to match the actual object boxes, it is necessary to enumerate all the possible positions and sizes of targets, leading to sample imbalance and a significant waste of computation.

In contrast, anchor-free methods abandon the predefined anchor approach and directly predict the key points of objects to determine their positions. These methods simplify the model structure and reduce computational load, offering faster detection speeds. CenterNet is a typical anchor-free object detection model \citep{duan2019centernet}. It locates and recognizes objects by detecting the central point of each object in the image and regressing from the center point to the target size. CenterNet does not require complicated anchor box setups and does not rely on complex candidate region proposal steps, making the model structure more straightforward and significantly reducing computational costs during training and inference. Additionally, CenterNet exhibits greater robustness in detecting small and densely distributed objects. And indeed, what we need is precisely the target's central point, which is where CenterNet excels. Therefore, we choose to use CenterNet as our model for the object detection stage.

\subsection{Image Classfication}

Image classification, which is also a fundamental task in computer vision, has undergone significant advancements over the past few decades. The evolution of this field has been marked by several key milestones and technological breakthroughs, particularly in the era of deep learning \citep{rawat2017deep}. The resurgence of CNNs began in 2012 with the introduction of AlexNet \citep{krizhevsky2012imagenet}. This deep CNN architecture achieved unprecedented accuracy in the ImageNet dataset \citep{deng2009imagenet}, significantly outperforming traditional methods. This watershed moment sparked a renewed interest in deep learning for image classification and initiated a period of rapid development in the field.

Following AlexNet, a series of increasingly sophisticated CNN architectures were proposed, each pushing the boundaries of image classification performance. Notable examples include VGGNet \citep{simonyan2014very}, GoogLeNet \citep{szegedy2015going}, and ResNet \citep{he2016deep}. These developments not only improved classification accuracy but also enhanced the efficiency and scalability of CNN models. The progress in CNN architectures was accompanied by advancements in optimization techniques, regularization methods, and data augmentation strategies, further boosting performance \citep{lu2020multiobjective}.

In recent years, the focus has shifted towards developing more efficient and compact models for real-world applications. This has led to the emergence of architectures like MobileNet and EfficientNet, which prioritize computational efficiency without significantly compromising accuracy \citep{howard2017mobilenets}. The rapid evolution of image classification techniques has had far-reaching implications across various domains.

ResNet, as one of the most reliable and widely used architectures for image classification tasks, has been proven effective across various applications due to its ability to mitigate the vanishing gradient problem through the introduction of residual connections. In this work, we also adopt ResNet as the backbone architecture for the second stage of the \textsc{DRAFTS} pipeline.

\section{Method}

\textsc{DRAFTS} identifies the arrival time and DM of signals within the data. This approach addresses the incompleteness, low operation efficiency, and dependence on manual inspection of a large number of false signals characteristic of traditional methods. The flowchart of our search pipeline is shown in Fig.~\ref{fig:search}.

Note that in Fig.~\ref{fig:radiodata} fourth column, the signal's S/N is highest and the width is narrowest when the data is de-dispersed with the correct DM value. As the DM value deviates from the optimum, the S/N of the signal decreases, and the width of the signal broadens. Hence, in the time-DM plot (Fig.~\ref{fig:radiodata} third column), bursts manifest the ``bow-tie" pattern. The coordinates at the center of the ``bow-tie" correspond to the burst's arrival time and DM value.

Accordingly, in this pipeline,
\begin{enumerate}
    \item We apply a range of DM values to de-disperse the data, transforming the original time-frequency data into time-DM data. During this process, we use \texttt{numba.cuda} \citep{lam2015numba} for acceleration. Tests have shown that \texttt{numba.cuda} can reduce the de-dispersion processing time on RTX 2070S to 1/1000 of the same processing time on Intel i7-10700K.
    \item Subsequently, the time-DM data is fed into a pre-trained object detection model (here CenterNet \citep{duan2019centernet}) to detect the signal's arrival time and DM value. At this step, instead of converting data into image files and reading it into the model for detection, we directly input the data stream, saving I/O time.
    \item Based on the arrival time and dispersion measure found through object detection, the signal is extracted from the original data, and a pre-trained classification model (here ResNet \citep{he2016deep}) is employed to determine the authenticity of the signal.
\end{enumerate}

The use of object detection ensures that the same signal is not repeatedly detected, and the occurrence of false signals is rare. Even if a false signal is detected, it will undergo secondary validation by the classification model, meaning that manual inspection is almost unnecessary, and search efficiency is greatly enhanced.

It is also worth mentioning that for follow-up observations of FRBs with known DM values, one could rely solely on the classification model for detection. We could first uniformly de-disperse the observation data according to its specific DM value and then segment the data, allowing the classification model to directly determine whether there are signals similar to those in Fig.~\ref{fig:radiodata} fourth column within the data slices.

\subsection{Data and Augmentation}

\paragraph{Object Detection}

To train our object detection model, we utilized a dataset comprising 2728 bursts detected by Five-hundred-meter Aperture Spherical radio Telescope (FAST) from FRB 20121102A \citep{2021Natur.598..267L} and FRB 20220912A \citep{2023ApJ...955..142Z}. %

We began by de-dispersing the original time-frequency data with dispersion measures ranging from $1\,\rm pc\,cm^{-3}$ to $1024\,\rm pc\,cm^{-3}$, with a step size of $1\,\rm pc\,cm^{-3}$, totaling 1024 DM values. During the process of de-dispersion, we divided the data into three frequency slices: low-frequency (1000-1250 MHz), high-frequency (1250-1500 MHz), and full-frequency (1000-1500 MHz). This division enhances the detection of narrow-band bursts, allowing us to capture signals that may only appear within a specific frequency range.

After segmenting the time-DM data, we performed manual labeling. The labels for files can be seen in Tab.~\ref{tab:lab}, where each row corresponds to a burst. The same file name may appear multiple times, indicating multiple bursts within the same file. The labels include the frequency slice, center point of the burst in the data, along with the half-width and half-height of the burst. For frequency slicing labels, 0 represents the full frequency, 1 denotes low frequency slice, and 2 indicates high frequency slice. A label of -1, -1, -1, -1 indicates that there was no burst in the file. 

\startlongtable
\begin{deluxetable*}{cccccc}
    \tablecaption{{\bf Example labels for data used in object detection.} The two rows with a gray background represent examples where multiple bursts are labeled in the same file, while the row with a yellow background shows an example where no bursts were found in the frequency slice of that file. \label{tab:lab}}
    \tablehead{
        \colhead{File Name} & \colhead{Frequency Slice} & \colhead{Time Center} & \colhead{DM Center} & \colhead{Time Width} & \colhead{DM Height}
    }
    \startdata
        0000.npy & 0 & 7743.7 & 564.66 & 7613.21 & 627.03 \\
        \rowcolor{brown!20}
        0000.npy & 1 & -1.0 & -1.0 & -1.0 & -1.0 \\
        0000.npy & 2 & 7766.73 & 554.11 & 7551.8 & 657.74 \\
        0001.npy & 0 & 628.03 & 552.19 & 221.21 & 602.09 \\
        0001.npy & 1 & 612.68 & 554.11 & 259.59 & 594.41 \\
        0001.npy & 2 & -1.0 & -1.0 & -1.0 & -1.0 \\
        0002.npy & 0 & 1134.65 & 551.23 & 1011.84 & 619.36 \\
        0002.npy & 1 & -1.0 & -1.0 & -1.0 & -1.0 \\
        0002.npy & 2 & 1142.33 & 552.19 & 973.46 & 625.11 \\
        0003.npy & 0 & 2347.46 & 548.35 & 1909.93 & 603.05 \\
        0003.npy & 1 & 2339.79 & 548.35 & 1963.66 & 592.49 \\
        0003.npy & 2 & -1.0 & -1.0 & -1.0 & -1.0 \\
        \rowcolor{gray!20}
        0004.npy & 0 & 5609.77 & 548.35 & 5348.79 & 591.53 \\
        \rowcolor{gray!20}
        0004.npy & 0 & 7567.16 & 558.91 & 7267.79 & 634.71 \\
        0004.npy & 1 & 5602.09 & 551.23 & 5287.38 & 592.49 \\
        0004.npy & 1 & 7582.51 & 553.15 & 7359.9 & 598.25 \\
        0004.npy & 2 & 5617.45 & 550.27 & 5533.01 & 590.57 \\
        0004.npy & 2 & 7620.89 & 552.19 & 7382.93 & 609.76 \\
        0005.npy & 0 & 3867.31 & 551.23 & 3690.77 & 608.8 \\
        0005.npy & 1 & -1.0 & -1.0 & -1.0 & -1.0 \\
        0005.npy & 2 & 3844.29 & 547.39 & 3721.47 & 611.68 \\
    \enddata
\end{deluxetable*}

\begin{figure*}[!htbp]
    \centering
    \includegraphics[width=\textwidth]{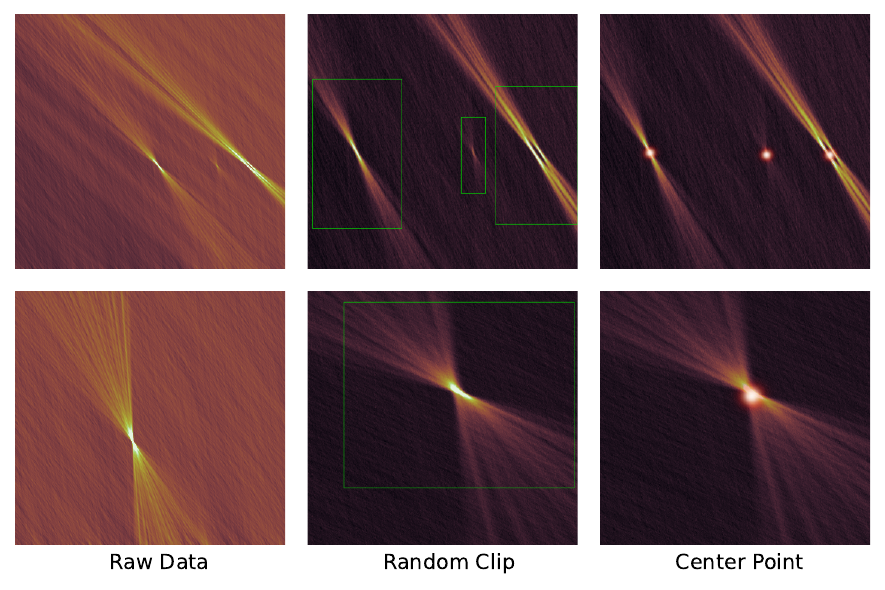}
    \caption{{\bf Data augmentation for object detection - Random Clipping.} The figure illustrates the process of random clipping applied to the input data. The first column displays the original input images, the second column shows the images after random cropping with green boxes indicating the ground truth labels, and the third column presents the center points after Gaussian scattering, which are used for CenterNet training.}
    \label{fig:augclip}
\end{figure*}

\begin{figure*}[!htbp]
    \centering
    \includegraphics[width=\textwidth]{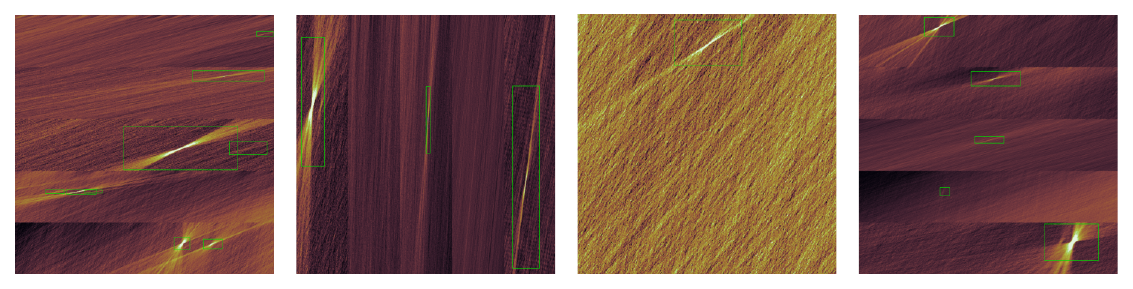}
    \caption{{\bf Data augmentation for object detection - Random Combination.} This figure demonstrates four examples of the random combination applied to the training data.}
    \label{fig:augcomb}
\end{figure*}

During training, we standardized the transformation of images to a size of $512 \times 512$. To ensure that the model's training results are robust, transferable, and capable of detecting signals of various shapes, we increased data variability by applying random cropping (as shown in Fig.~\ref{fig:augclip}), and and random combining of 1-5 files  (as shown in Fig.~\ref{fig:augcomb}).

\paragraph{Binary Classification}

\begin{figure*}[!htbp]
    \centering
    \includegraphics[width=\textwidth]{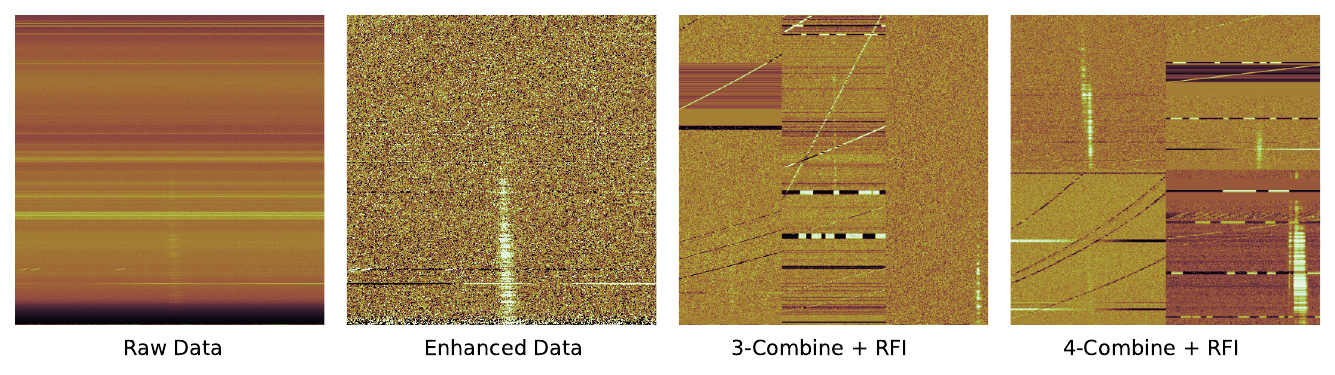}
    \caption{{\bf Data augmentation for binary classification.} The first column is the input data, the second column is the data after enhancement, and the third and fourth columns are the data after adding interference and randomly combining.}
    \label{fig:augmentation2}
\end{figure*}

To train our classification model, we also utilized the dataset of 2728 bursts detected by FAST. This time we rely on the time-frequency data after de-dispersion. We uniformly process the original data using a DM value of $565\,\rm pc\ cm^{-3}$ for FRB 20121102A and $220\,\rm pc\ cm^{-3}$ for FRB 20220912A to de-disperse, and isolate the data segments containing bursts for training purposes. We then perform data augmentation to increase the diversity of the data, as shown in Fig.~\ref{fig:augmentation2}. The augmentation process includes the following steps:

\begin{enumerate}
    \item Each frequency channel of the data is divided by its mean value, and the data's numerical range is confined to its $10\%-90\%$ dynamic range to boost the S/N as much as possible.
    \item Randomly combine 1-5 images into a single image to increase the robustness and generalizability of the model. Fig.~\ref{fig:augmentation2} illustrates the case of merging three and four images; if there are four images, they are randomly combined horizontally, vertically, or in a $2\times2$ configuration. In other instances, images are randomly concatenated along either the horizontal or vertical axis.
    \item Artificial interference is introduced into the data using random numbers to avoid the inability of a limited dataset to cover as many RFI morphologies as possible, which would result in the model failing to generalize to new data or data from other telescopes. The interferences we add include broadband interference with a DM of $0\,\rm pc\ cm^{-3}$, narrowband interference that varies over time or is invariant with time, as well as some random scatter points.
    \item Randomly rotate and flip the images.
\end{enumerate}

All training data are finally transformed to a size of $512 \times 512$.

\subsection{Training and Inference}

For object detection, we constructed a minimal implementation of CenterNet based on PyTorch \citep{paszke2019pytorch}. The input is data with dimensions of $1\times512\times512$. The chosen backbone for the network is ResNet18, with a comparative use of ResNet50. The output is an array with a size of $5\times128\times128$, where the first channel represents the center point, the subsequent two channels denote width and height, and the final two channels correspond to the offset of the center point. Consequently, the loss function of CenterNet is comprised of these three components.

\begin{equation}
    \mathcal{L} = \mathcal{L}_{\rm center} + \lambda_{\rm size}\mathcal{L}_{\rm size} + \lambda_{\rm offset}\mathcal{L}_{\rm offset}
\end{equation}
where $\mathcal{L}_{\rm center}$ is the focal loss for the center point, $\mathcal{L}_{\rm size}$ is the smooth L1 loss for the width and height, and $\mathcal{L}_{\rm offset}$ is the smooth L1 loss for the offset.

Owing to the high precision required for the center point localization, we set the weight $\lambda_{\rm size}=0.1$, which have a lesser association with the center point, and $\lambda_{\rm offset}=1.0$. We employed the Adam optimizer \citep{2014arXiv1412.6980K} along with a CosineLRScheduler strategy \citep{2016arXiv160803983L} for the learning rate decay. The training parameters are presented in Tab.~\ref{tab:params}.

\begin{deluxetable}{cc}
    \tablecaption{{\bf Training parameters for CenterNet.} \label{tab:params}}
    \tablehead{\colhead{Parameter} & \colhead{Value}}
    \startdata
        Batch Size         & 4    \\
        Learning Rate (LR) & 1e-3 \\
        Num Epochs         & 100  \\
        Warmup Epochs      & 5    \\
        Initial Warmup LR  & 1e-5 \\
        Minimum LR         & 1e-5 \\
    \enddata
\end{deluxetable}

For classification model, we utilized the ResNet18 architecture as our classification model, with a comparative use of ResNet50. The model's input is an array of size $1\times512\times512$, with the output being a probability ranging from $0$ to $1$, activated by the sigmoid function. The loss function employed is BCELoss, accompanied by the Adam optimizer along with a CosineLRScheduler strategy for the learning rate decay. Training parameters are analogous to those used for CenterNet, with the exception of the batch size, which is set to 32.

\section{Experiments}

We evaluate our model's performance utilizing the independent dataset from ``FAST dataset for Fast Radio bursts EXploration" (FAST-FREX) \citep{fast-fex}. This dataset comprises 600 burst samples originating from three distinct FRBs. Each burst is stored within a \textsc{fits} file, which contains approximately 6.04 seconds of data, along with the best DM value for each burst.

We deploy both the object detection model and binary classification model to search for bursts within the dataset, contrasting these techniques with \textsc{presto} as a baseline. All experiments were carried out on a computer with Intel i7-10700K, RTX 2070S, and 32GB memory.

For the \textsc{presto} search, we adhere to the standard workflow involving the utilization of \texttt{rfifind}, \texttt{prepsubband}, and \texttt{single\_pulse\_search} commands, sequentially conducting RFI mitigation, generating time series data through de-dispersion at a set of DM values, and searching for potential high S/N events within the time series data. Owing to the slow processing times of this tool, we limit our de-dispersion to 100 DM values per file, centered around the burst's optimal DM. For instance, if a burst has a dispersion measure of $550\,\rm pc\,cm^{-3}$, we select DM values ranging from $500\,\rm pc\,cm^{-3}$ to $599\,\rm pc\,cm^{-3}$, with a step size of $1\,\rm pc\,cm^{-3}$. Despite this restriction, \textsc{presto} still requires an average of 120 seconds to process a 6-second file. After generating time series for 100 DM values, we conduct a single pulse search and compile the results. We record the number of detected bursts, missed bursts, false positives, and the total count of duplicate detections for the same burst at different DM values at S/N thresholds of 3, 5, and 7.

For the search using our object detection model, we employ \textsc{astropy} \citep{price2022astropy} to read the time-frequency data saved in \textsc{fits} files. We process de-dispersion for DM values from $1\,\rm pc\,cm^{-3}$ to $1024\,\rm pc\,cm^{-3}$ with a step of $1\,\rm pc\,cm^{-3}$ using \texttt{numba.cuda}, converting the original time-frequency data to time-DM data, and input this into the pretrained centernet model for prediction. Based on the predicted bounding box centers, we determine the arrival times and DM values of the detected bursts and tally the number of bursts found, missed, and falsely identified. The model applies non-maximum suppression (NMS) during prediction, thereby eliminating multiple counts of the same burst, and thus, we do not record duplicates. The CenterNet model, when using ResNet18 as the backbone, processes a 6-second file in roughly 4.5 seconds, and with ResNet50 as the backbone, necessitates about 4.7 seconds, both including the time for file reading, writing and de-dispersion.

Regarding the search performed by our binary classification model, we similarly utilize \textsc{astropy} to read time-frequency data from \textsc{fits} files. We perform uniform de-dispersion based on the DM values for the three FRB sources, then partition the data into non-overlapping segments, resize these to $512\times512$, and feed them into the trained ResNet model to determine the presence of bursts within the data segments. We also tabulate the number of bursts found, missed, and falsely identified. The process leveraging ResNet18 and ResNet50 are both around 1.2 seconds to handle a 6-second file, both including the time for file reading, writing and de-dispersion. The results are shown in Tab.~\ref{tab:performatnce}.

\begin{deluxetable*}{ccccccccc}
\tablecaption{Performance comparison of different methods.\label{tab:performatnce}}
    \tablehead{
        \colhead{Method} & \colhead{Threshold} & \colhead{TP} & \colhead{FP} & \colhead{Missed} & \colhead{Duplicates} & \colhead{Precision} & \colhead{Recall} & \colhead{Time (s)}
    }
    \startdata
        Presto       & S/N = 3   & 520 & 10663950 & 80     & 43044      & 0.0049\%  & 86.7\% & $\sim 120$ \\
        Presto       & S/N = 5   & 513 & 17406    & 87     & 40818      & 2.8\%     & 85.5\% & -          \\
        Presto       & S/N = 7   & 477 & 4488     & 123    & 25402      & 9.6\%     & 79.5\% & -          \\
        \hline
        CenterNet-18 & 0.5       & 580 & 23       & 20     & -          & 96.2\%    & 96.7\% & 4.51       \\
        CenterNet-50 & 0.5       & 578 & 20       & 22     & -          & 96.7\%    & 96.3\% & 4.67       \\
        \hline
        ResNet-18    & 0.5       & 600 &  1       & 0      & -          & 99.8\%    & 100\%  & 1.16       \\
        ResNet-50    & 0.5       & 600 &  1       & 0      & -          & 99.8\%    & 100\%  & 1.23       \\
    \enddata
\end{deluxetable*}

As the benchmark, \textsc{presto} exhibits an increasing aptitude to recall signals as the S/N threshold is lowered. However, the increment in genuine signals from a S/N drop from 5 to 3 is markedly less than that experienced in a reduction from 7 to 5, while the number of spurious signals has surged considerably. The optimal recall rate peaked at 86.7\%. Notably, our application of \textsc{presto} was confined to processing only 100 DM values. If we were to extend the de-dispersion to 1024 DM values, akin to CenterNet, the computational burden would escalate by an order of magnitude, implying a tenfold increase in data processing time. This would lead to an increase in false signals without a commensurate rise in real ones.

\begin{figure*}[!htbp]
    \centering
    \includegraphics[width=\textwidth]{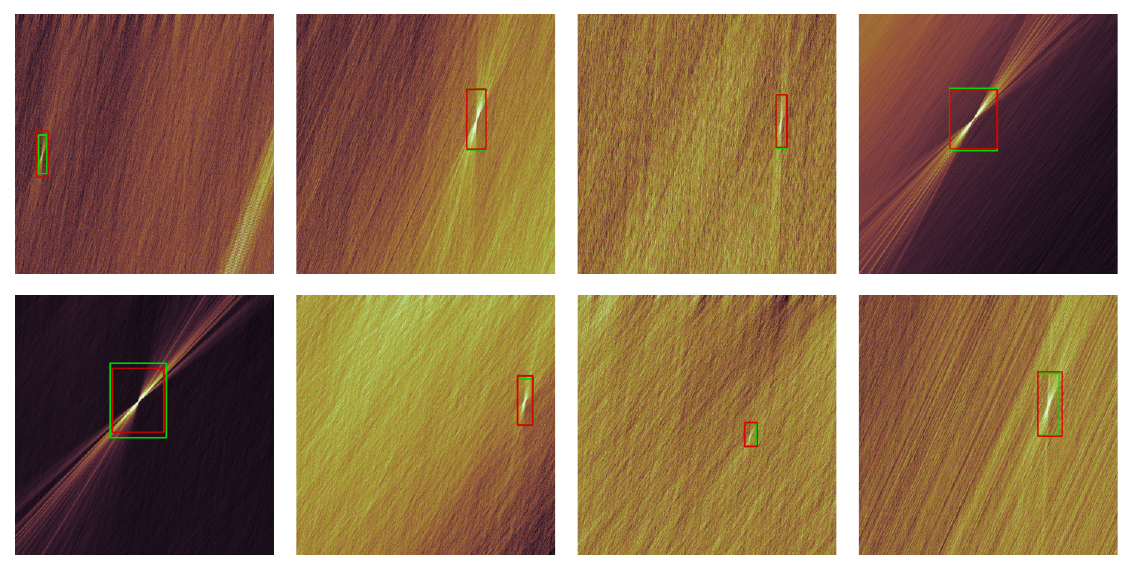}
    \caption{{\bf Examples of detected signals using CenterNet.} The green boxes indicating the ground truth labels, and the red boxes are the predicted labels.}
    \label{fig:detect}
\end{figure*}

In stark contrast, the object detection and classification models both approach a near-perfect recall rate, also upholding exceedingly high precision, thus demonstrating efficiencies far exceeding those of traditional search methods. Fig.~\ref{fig:detect} illustrates some signal examples inferred through CenterNet, which highlights the model's resilience; it adeptly discerns the `bow-tie' signature characteristic of FRBs, despite the significant scale variances occurring due to the disparities in signal strength of FRB events, spanning several orders of magnitude. The model's robustness is showcased as it remains functional even under the challenging conditions posed by such intense variations in the magnitude of the signal bursts.

To demonstrate the effectiveness of our classification model, specifically its ability to accurately identify the presence of a burst in an image and thus classify the data as true, we utilize Grad-CAM++ \citep{chattopadhay2018grad} to visualize the regions in the image that significantly influence the network's decision. Grad-CAM (Gradient-weighted Class Activation Mapping) is a technique that provides visual explanations for decisions made by convolutional neural networks (CNNs) \citep{selvaraju2017grad}. It generates class-specific activation maps using gradient information, highlighting the important regions in the input image that contribute to the model's prediction. The process involves computing gradients of the target class score concerning feature maps of the last convolutional layer, which are then globally averaged to obtain the importance weights. Grad-CAM++ is an enhanced version of Grad-CAM that offers more precise and detailed visual explanations, especially when multiple objects are present in the image \citep{chattopadhay2018grad}.

\begin{figure*}[!htbp]
    \centering
    \includegraphics[width=\textwidth]{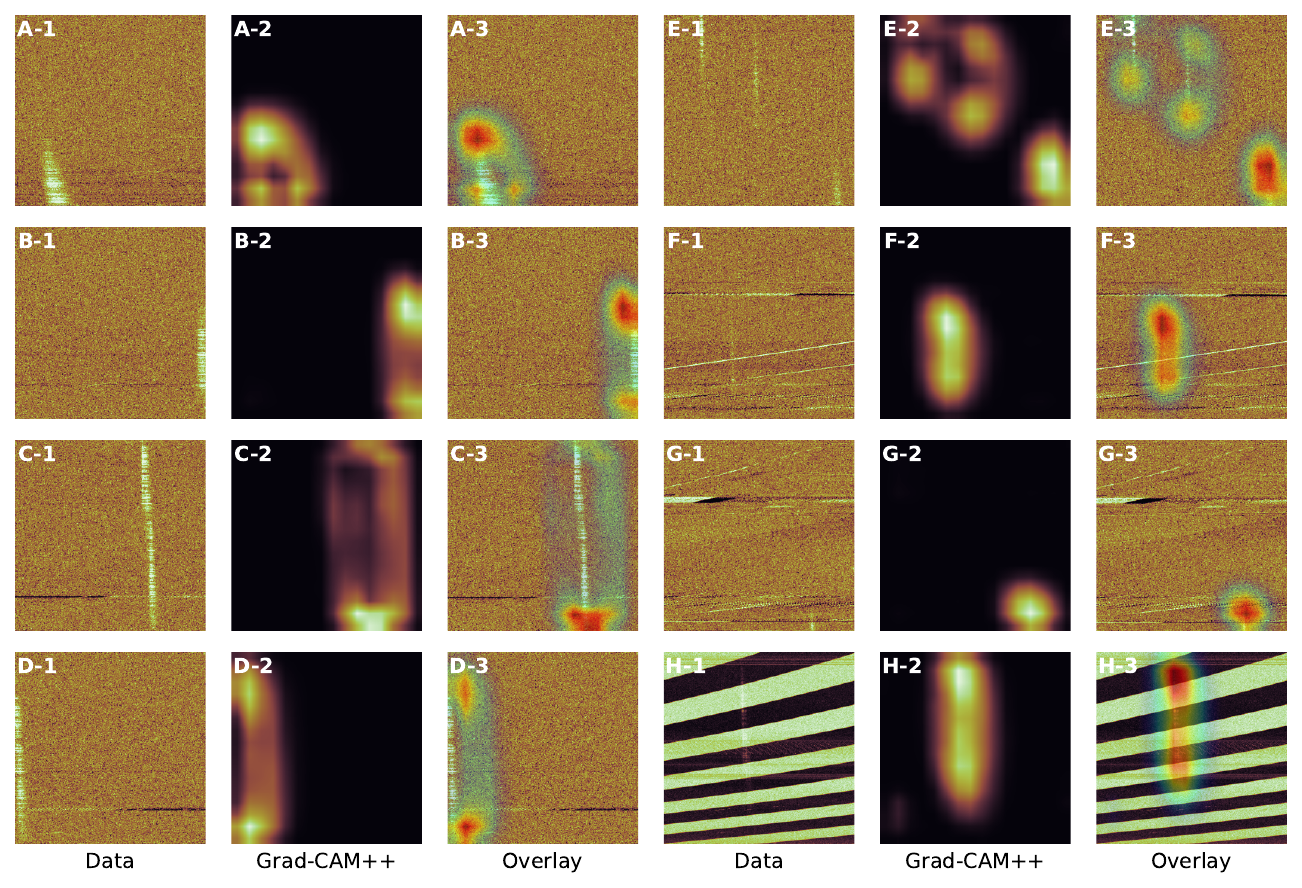}
    \caption{{\bf Examples of detected signals using ResNet.} A-H are 8 instances of data classified as true by the classification model, where 1, 2, and 3 represent the original data, the regions in the image that significantly influence the network's decision visualized using Grad-CAM++, and the result of superimposing these regions on the original data, respectively.}
    \label{fig:class}
\end{figure*}

Fig.~\ref{fig:class} shows the results of Grad-CAM++ visualization, containing eight instances of data classified as true by the model. Columns in Fig.~\ref{fig:class} represents the original data, the visualization of the critical regions influencing the network's decision using Grad-CAM++, and the result of superimposing these regions on the original data. It is evident that the model's attention is indeed focused on the location of the burst. Even when the burst is at the edge (Fig.~\ref{fig:class} B and D), or there is strong interference in the data (Fig.~\ref{fig:class} F and G), even in the noise injected data  (Fig.~\ref{fig:class} H), the model can still make accurate and effective judgments.

Additionally, we observe that ResNet-50 performs similarly to ResNet-18 in both object detection and classification frameworks. Therefore, to balance computational efficiency with model performance, we choose ResNet-18 as the backbone network for both CenterNet and the classification model.

\section{Applications to FRB 20190520B}

To further validate the capability of \textsc{DRAFTS}, we applied the model to the 2020 FAST observation data of FRB 20190520B. In \cite{2022Natur.606..873N}, the discovery of FRB 20190520B was reported, and multiple observations of this FRB were conducted between April 24, 2020, and September 19, 2020. During this period, a total of 75 bursts were detected using \textsc{Heimdall}, with an estimated event rate of $4.5^{+1.9}_{-1.5}\,\rm hr^{-1}$. The details of each observation, including the start time and duration of each session, are listed in Tab.~\ref{tab:obinfo}.

\begin{deluxetable*}{ccccccc}
    \tablecaption{{\bf FAST observation details of FRB 20190520B.} \label{tab:obinfo}}
    \tablehead{
        Date & Start MJD & Duration & Old Number & Add Number & Total Number & Average Burst Rate \\
        & (topocentric) & (minutes) & \citep{2022Natur.606..873N} & (This Paper) & & (hr$^{-1}$)
    }
    \startdata
    20200424 & 58963.742361111 & 108.0 & 2 & 4 & 6 & 3.3 \\
    \hline
    20200522 & 58991.664768519 & 118.0 & 13 & 9 & 22 & 11.2 \\
    \hline
    \multirow{2}{*}{20200730} & 59060.475694444 & 16.0 & 1 & 3 & 4 & 15.0 \\
    & 59060.494490741 & 91.9 & 2 & 15 & 17 & 11.1 \\
    \hline
    20200731 & 59061.490902778 & 83.1 & 12 & 16 & 28 & 20.2 \\
    \hline
    \multirow{2}{*}{20200806} & 59067.462800926 & 14.1 & 1 & 1 & 2 & 8.5 \\
    & 59067.479467593 & 79.6 & 5 & 33 & 38 & 28.6 \\
    \hline
    \multirow{2}{*}{20200808} & 59069.451388889 & 10.4 & 0 & 1 & 1 & 5.8 \\
    & 59069.465277778 & 90.0 & 3 & 27 & 30 & 20.0 \\
    \hline
    \multirow{2}{*}{20200810} & 59071.445833333 & 14.5 & 0 & 2 & 2 & 8.3 \\
    & 59071.462141204 & 87.0 & 3 & 8 & 11 & 7.6 \\
    \hline
    \multirow{2}{*}{20200812} & 59073.441342593 & 10.0 & 0 & 1 & 1 & 6.0 \\
    & 59073.452627315 & 93.6 & 3 & 13 & 16 & 10.3 \\
    \hline
    \multirow{2}{*}{20200814} & 59075.437835648 & 10.9 & 0 & 2 & 2 & 11.0 \\
    & 59075.451944444 & 69.5 & 5 & 17 & 22 & 19.0 \\
    \hline
    \multirow{2}{*}{20200816} & 59077.430555556 & 10.0 & 0 & 4 & 4 & 23.9 \\
    & 59077.444629630 & 90.2 & 20 & 7 & 27 & 18.0 \\
    \hline
    20200828 & 59089.413194444 & 60.0 & 2 & 10 & 12 & 12.0 \\
    \hline
    \multirow{2}{*}{20200919} & 59111.346608796 & 6.7 & 0 & 0 & 0 & 0.0 \\
    & 59111.356840278 & 36.4 & 3 & 10 & 13 & 21.4 \\
    \bottomrule
    \multicolumn{2}{c}{\textbf{Total}} & 1100 & 75 & 183 & 258 & 14.1 \\
    \enddata
\end{deluxetable*}

We re-searched this data using \textsc{DRAFTS} and detected all 75 previously discovered bursts, along with an additional 183 new bursts, more than doubling the original number of detected bursts and bringing the maximum event rate during this period to $28.6\,\rm hr^{-1}$. The arrival times of these 183 bursts are listed in Tab.~\ref{tab:mjd}, and their dynamic spectra are shown in Fig.~\ref{fig:newburst}.

\begin{figure}[!htbp]
    \centering
    \includegraphics[width=0.48\textwidth]{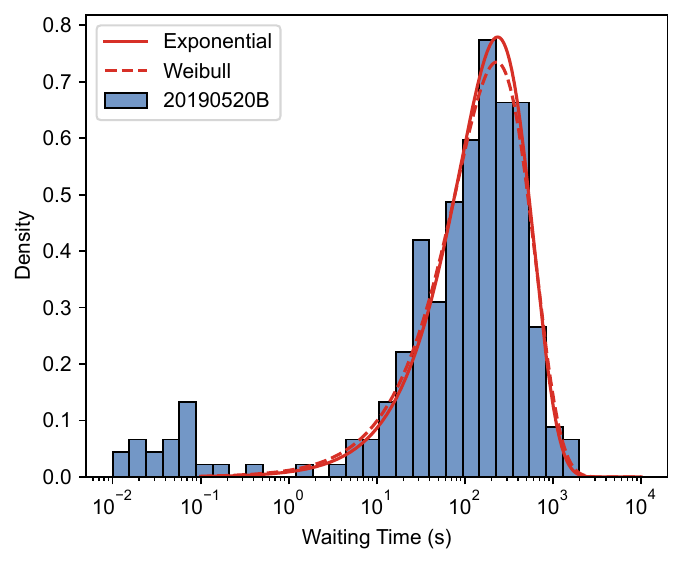}
    \caption{{\bf Waiting time distribution of FRB 20190520B.} The red solid line represents the exponential distribution fit, the red dashed line represents the Weibull distribution fit, and the blue bars represent the waiting time distribution of FRB 20190520B.}
    \label{fig:wt}
\end{figure}

We further estimated the event rate during this period using waiting time analysis. Fig.~\ref{fig:wt} shows the waiting time distribution of FRB 20190520B, including the newly detected bursts. Although the left peak is not very prominent, the overall distribution still exhibits a bimodal pattern, with the right peak corresponding to the FRB's active phase \citep{2024SciBu..69.1020Z}. We fitted the waiting time distribution for intervals longer than 1 second using both exponential and Weibull distributions. For the fitting, we employed the \textsc{EMCEE} package to perform maximum likelihood estimation of the fitting parameters. For the exponential distribution, we defined the likelihood function as

\begin{equation}
    L(\lambda | t) = \sum_i \log \left(\lambda e^{-\lambda t}\right)
\end{equation}
where \( t \) represents the waiting time, and \( 1/\lambda \) corresponds to the event rate.

For the Weibull distribution, the likelihood function is defined as

\begin{equation}
    L(\lambda, k | t) = \sum_i \log \left[\frac{k}{\lambda} \left( \frac{t}{\lambda} \right)^{k-1} e^{-(t/\lambda)^k}\right]
\end{equation}
where \( k \) is the shape parameter and \( \lambda \) is the scale parameter. The event rate corresponds to the reciprocal of the Weibull distribution's expected value, i.e., \( 1/[\lambda\Gamma(1+1/k)] \).

As shown in Fig.~\ref{fig:wt}, both distributions provide a good fit to the waiting time distribution of FRB 20190520B. The event rate estimated from the exponential fit is \( 15.29^{+1.02}_{-0.99}\,\rm hr^{-1} \), while the Weibull fit yields an event rate of \( 15.29^{+1.11}_{-1.07}\,\rm hr^{-1} \) with a shape parameter \( k=0.94^{+0.05}_{-0.05} \). The two estimates are nearly identical and close to the average event rate estimated in Tab.~\ref{tab:obinfo}. Furthermore, for the Weibull distribution, \( k \) is close to 1. In fact, when \( k = 1 \), the Weibull distribution reduces to an exponential distribution. Therefore, we can reasonably conclude that the bursts of FRB 20190520B can be considered as samples from a Poisson process with a constant event rate of \( 15.29\,\rm hr^{-1} \), indicating that this FRB is far more active than previously thought.

Testing on the complete set of real observation data from FAST further confirms that \textsc{DRAFTS} is effective in detecting FRBs. The number of bursts identified by \textsc{DRAFTS} is more than three times higher than that detected by \textsc{Heimdall}, which holds substantial importance for subsequent statistical analyses.

\section{Limitations and Conclusions}

\begin{figure*}[!htbp]
    \centering
    \includegraphics[width=\textwidth]{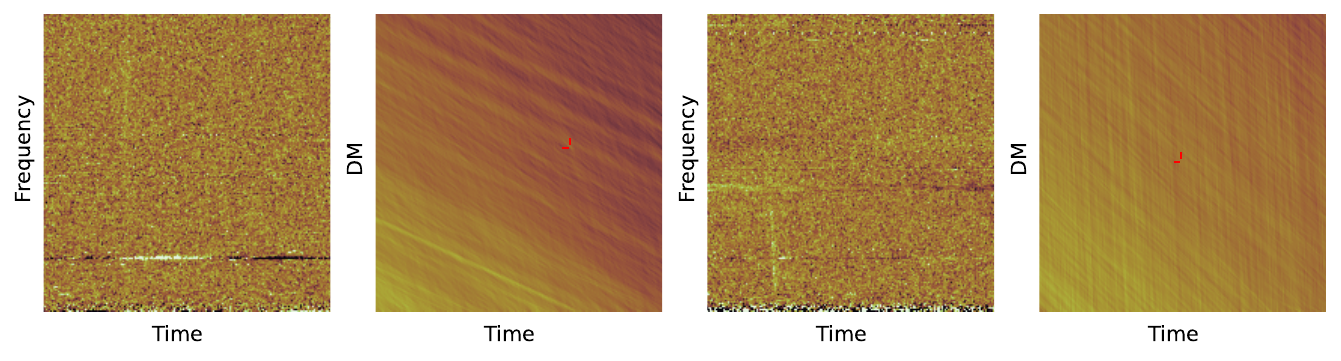}
    \caption{{\bf Two examples of omitted signals.} The first and third columns show the time-frequency plots of two bursts, the second and fourth columns display the time-DM plots. The expected bursts' location in time-DM plots are marked by red lines.}
    \label{fig:omit}
\end{figure*}

As shown in Tab.~\ref{tab:performatnce}, the classification model manifests a recall rate of 100\%, while CenterNet's recall rate is marginally below this benchmark. Fig.~\ref{fig:omit} embodies the time-frequency and time-dispersion plots for two specific burst events identified by the classification model but overlooked by the object detection model. It is evident from the plots that the signals from these bursts are exceedingly weak, and the `bow-tie' feature within the time-dispersion plots is ambiguously defined - practically invisible to the naked eye. Thus, it is understandable that these signals eluded detection by the object detection model.

This calls for the advancement of our methodology concerning the conversion of raw time-frequency data to time-dispersion data, with a special emphasis on improving the visibility of weaker signals. Enhancing this process would potentially mitigate the issue of non-detection in target models and lead to a more reliable and efficient identification of transient astronomical events driven by faint signals.

In conclusion, we have developed a comprehensive training dataset of large-scale, real-world data for FRB search and created \textsc{DRAFTS}, an advanced tool that integrates object detection with binary classification to identify FRBs in radio data. Our experiments reveal that \textsc{DRAFTS} significantly outperforms traditional methods in detection speed, accuracy and completeness. This pipeline not only facilitates real-time FRB detection but also holds potential for application in other radio transient searches.

Future work will focus on enhancing the visibility of faint signals within time-dispersion plots, improving the overall efficiency of the search pipeline and add more data in the training set. The deployment of \textsc{DRAFTS} represents a groundbreaking and reliable approach, offering substantial benefits for observational campaigns. By employing this workflow, we anticipate a notable acceleration in the detection of radio transients, which will in turn drive a deeper understanding of the physical mechanisms behind these extreme cosmic events. Furthermore, \textsc{DRAFTS} provides a powerful impetus and essential tools for exploring the uncharted territories of the universe, ultimately contributing to our broader knowledge of cosmic phenomena.\\\\

This work was supported by the National Natural Science Foundation of China (NSFC) grant No. 11988101, 12203045 and by Key Research Project of Zhejiang Lab No. 2021PE0AC03. Di Li is a New Cornerstone Investigator. Pei Wang is supported by NSFC grant No. 12041303, the CAS Youth Interdisciplinary Team, the Youth Innovation Promotion Association CAS (id. 2021055), and the Cultivation Project for FAST Scientific Payoff and Research Achievement of CAMS-CAS. Chen-Hui Niu is supported by NSFC grant No. 12203069, 12041302, 12203045, the National SKA Program of China/2022SKA0130100, the Office Leading Group for Cyberspace Affairs, CAS (No. CAS-WX2023PY-0102), and the CAS Youth Interdisciplinary Team and the Foundation of Guizhou Provincial Education Department for Grants No. KY(2023)059.

\software{astropy \citep{price2022astropy}, numba \citep{lam2015numba}, Pytorch \citep{paszke2019pytorch}}

\clearpage
\appendix

\section{Arrival Times of Detected Bursts from FRB 20190520B}

\startlongtable
\begin{deluxetable}{ccccccccc}
    \tablecaption{{\bf MJD of arrival times for FRB 20190520B.} \label{tab:mjd}}
    \setlength\tabcolsep{3pt}
    \tablehead{
        Burst ID & MJD$^a$ & Date$^b$ & Burst ID & MJD$^a$ & Date$^b$ & Burst ID & MJD$^a$ & Date$^b$
    }
    \startdata
    \multicolumn{9}{l}{Newly detected bursts using \textsc{DRAFTS}.} \\
    \hline
    B001 & 58963.753958080 & 20200424 & B062 & 59067.502978234 & 20200806-2 & B123 & 59073.458422907 & 20200812-2 \\
    B002 & 58963.770184740 & 20200424 & B063 & 59067.503231597 & 20200806-2 & B124 & 59073.458586656 & 20200812-2 \\
    B003 & 58963.770865928 & 20200424 & B064 & 59067.504781097 & 20200806-2 & B125 & 59073.461018945 & 20200812-2 \\
    B004 & 58963.808134276 & 20200424 & B065 & 59067.505167148 & 20200806-2 & B126 & 59073.464894870 & 20200812-2 \\
    B005 & 58991.676149829 & 20200522 & B066 & 59067.506972167 & 20200806-2 & B127 & 59073.471089860 & 20200812-2 \\
    B006 & 58991.681677339 & 20200522 & B067 & 59067.511287838 & 20200806-2 & B128 & 59073.474148791 & 20200812-2 \\
    B007 & 58991.687041934 & 20200522 & B068 & 59067.511635139 & 20200806-2 & B129 & 59073.483559837 & 20200812-2 \\
    B008 & 58991.687858426 & 20200522 & B069 & 59067.513425862 & 20200806-2 & B130 & 59073.486190472 & 20200812-2 \\
    B009 & 58991.696814811 & 20200522 & B070 & 59067.518125365 & 20200806-2 & B131 & 59073.490224233 & 20200812-2 \\
    B010 & 58991.701210940 & 20200522 & B071 & 59067.518575780 & 20200806-2 & B132 & 59073.505812649 & 20200812-2 \\
    B011 & 58991.714680346 & 20200522 & B072 & 59067.518906594 & 20200806-2 & B133 & 59073.514778195 & 20200812-2 \\
    B012 & 58991.733097018 & 20200522 & B073 & 59067.522778481 & 20200806-2 & B134 & 59075.442356225 & 20200814-1 \\
    B013 & 58991.739498157 & 20200522 & B074 & 59067.525342925 & 20200806-2 & B135 & 59075.443059359 & 20200814-1 \\
    B014 & 59060.480277680 & 20200730-1 & B075 & 59067.525688371 & 20200806-2 & B136 & 59075.458608135 & 20200814-2 \\
    B015 & 59060.481182843 & 20200730-1 & B076 & 59067.525748592 & 20200806-2 & B137 & 59075.460956536 & 20200814-2 \\
    B016 & 59060.483628231 & 20200730-1 & B077 & 59067.527927305 & 20200806-2 & B138 & 59075.464442699 & 20200814-2 \\
    B017 & 59060.500614681 & 20200730-2 & B078 & 59067.527976492 & 20200806-2 & B139 & 59075.467864833 & 20200814-2 \\
    B018 & 59060.528391399 & 20200730-2 & B079 & 59067.527977138 & 20200806-2 & B140 & 59075.471004584 & 20200814-2 \\
    B019 & 59060.528752528 & 20200730-2 & B080 & 59067.528148897 & 20200806-2 & B141 & 59075.471743866 & 20200814-2 \\
    B020 & 59060.529584944 & 20200730-2 & B081 & 59067.531950533 & 20200806-2 & B142 & 59075.471744903 & 20200814-2 \\
    B021 & 59060.530094435 & 20200730-2 & B082 & 59069.452864158 & 20200808-1 & B143 & 59075.472468242 & 20200814-2 \\
    B022 & 59060.536139806 & 20200730-2 & B083 & 59069.465588641 & 20200808-2 & B144 & 59075.477403204 & 20200814-2 \\
    B023 & 59060.536711714 & 20200730-2 & B084 & 59069.466031951 & 20200808-2 & B145 & 59075.479022054 & 20200814-2 \\
    B024 & 59060.542483004 & 20200730-2 & B085 & 59069.466841138 & 20200808-2 & B146 & 59075.479137089 & 20200814-2 \\
    B025 & 59060.544197134 & 20200730-2 & B086 & 59069.472180281 & 20200808-2 & B147 & 59075.481013426 & 20200814-2 \\
    B026 & 59060.546182273 & 20200730-2 & B087 & 59069.472342565 & 20200808-2 & B148 & 59075.482251647 & 20200814-2 \\
    B027 & 59060.546973407 & 20200730-2 & B088 & 59069.476379684 & 20200808-2 & B149 & 59075.486885993 & 20200814-2 \\
    B028 & 59060.551389491 & 20200730-2 & B089 & 59069.480358138 & 20200808-2 & B150 & 59075.487163115 & 20200814-2 \\
    B029 & 59060.552381810 & 20200730-2 & B090 & 59069.480654383 & 20200808-2 & B151 & 59075.488591853 & 20200814-2 \\
    B030 & 59060.552999128 & 20200730-2 & B091 & 59069.481851782 & 20200808-2 & B152 & 59075.489545069 & 20200814-2 \\
    B031 & 59060.553319676 & 20200730-2 & B092 & 59069.482092574 & 20200808-2 & B153 & 59077.431731064 & 20200816-1 \\
    B032 & 59061.497093144 & 20200731 & B093 & 59069.486802716 & 20200808-2 & B154 & 59077.433249180 & 20200816-1 \\
    B033 & 59061.500199528 & 20200731 & B094 & 59069.488261607 & 20200808-2 & B155 & 59077.434196579 & 20200816-1 \\
    B034 & 59061.503304566 & 20200731 & B095 & 59069.493983403 & 20200808-2 & B156 & 59077.435368080 & 20200816-1 \\
    B035 & 59061.503605827 & 20200731 & B096 & 59069.496322401 & 20200808-2 & B157 & 59077.456186297 & 20200816-2 \\
    B036 & 59061.505380004 & 20200731 & B097 & 59069.504392562 & 20200808-2 & B158 & 59077.460669274 & 20200816-2 \\
    B037 & 59061.507535918 & 20200731 & B098 & 59069.504842690 & 20200808-2 & B159 & 59077.466908013 & 20200816-2 \\
    B038 & 59061.510490722 & 20200731 & B099 & 59069.504963898 & 20200808-2 & B160 & 59077.473437142 & 20200816-2 \\
    B039 & 59061.510702601 & 20200731 & B100 & 59069.505030985 & 20200808-2 & B161 & 59077.476236056 & 20200816-2 \\
    B040 & 59061.513361812 & 20200731 & B101 & 59069.508373735 & 20200808-2 & B162 & 59077.477652805 & 20200816-2 \\
    B041 & 59061.518917368 & 20200731 & B102 & 59069.510679931 & 20200808-2 & B163 & 59077.480078652 & 20200816-2 \\
    B042 & 59061.525294977 & 20200731 & B103 & 59069.513873763 & 20200808-2 & B164 & 59089.422578365 & 20200828 \\
    B043 & 59061.531004139 & 20200731 & B104 & 59069.514277920 & 20200808-2 & B165 & 59089.426408841 & 20200828 \\
    B044 & 59061.532807852 & 20200731 & B105 & 59069.518708503 & 20200808-2 & B166 & 59089.428926655 & 20200828 \\
    B045 & 59061.542664556 & 20200731 & B106 & 59069.518777351 & 20200808-2 & B167 & 59089.434616349 & 20200828 \\
    B046 & 59061.547473763 & 20200731 & B107 & 59069.519294932 & 20200808-2 & B168 & 59089.438048012 & 20200828 \\
    B047 & 59061.548427662 & 20200731 & B108 & 59069.522911272 & 20200808-2 & B169 & 59089.443650739 & 20200828 \\
    B048 & 59067.463647943 & 20200806-1 & B109 & 59069.525875712 & 20200808-2 & B170 & 59089.443845199 & 20200828 \\
    B049 & 59067.480205948 & 20200806-2 & B110 & 59071.451254334 & 20200810-1 & B171 & 59089.451226974 & 20200828 \\
    B050 & 59067.481172292 & 20200806-2 & B111 & 59071.453382840 & 20200810-1 & B172 & 59089.451227866 & 20200828 \\
    B051 & 59067.482660405 & 20200806-2 & B112 & 59071.469047453 & 20200810-2 & B173 & 59089.454059447 & 20200828 \\
    B052 & 59067.484752349 & 20200806-2 & B113 & 59071.470776754 & 20200810-2 & B174 & 59111.359668520 & 20200919-2 \\
    B053 & 59067.485940012 & 20200806-2 & B114 & 59071.478912028 & 20200810-2 & B175 & 59111.362133525 & 20200919-2 \\
    B054 & 59067.488051465 & 20200806-2 & B115 & 59071.487867787 & 20200810-2 & B176 & 59111.363547425 & 20200919-2 \\
    B055 & 59067.489458573 & 20200806-2 & B116 & 59071.496511074 & 20200810-2 & B177 & 59111.364054532 & 20200919-2 \\
    B056 & 59067.491462488 & 20200806-2 & B117 & 59071.499713051 & 20200810-2 & B178 & 59111.365293865 & 20200919-2 \\
    B057 & 59067.493311787 & 20200806-2 & B118 & 59071.500538800 & 20200810-2 & B179 & 59111.366311841 & 20200919-2 \\
    B058 & 59067.493315628 & 20200806-2 & B119 & 59071.516139769 & 20200810-2 & B180 & 59111.368587448 & 20200919-2 \\
    B059 & 59067.496107313 & 20200806-2 & B120 & 59073.445652004 & 20200812-1 & B181 & 59111.373554797 & 20200919-2 \\
    B060 & 59067.496636351 & 20200806-2 & B121 & 59073.454481887 & 20200812-2 & B182 & 59111.377382763 & 20200919-2 \\
    B061 & 59067.500936084 & 20200806-2 & B122 & 59073.457517769 & 20200812-2 & B183 & 59111.379347916 & 20200919-2 \\
    \hline
    \multicolumn{9}{l}{Bursts detected using \textsc{Heimdall} in \cite{2022Natur.606..873N}} \\
    \hline
    B001 & 58963.760965927 & 20200424 & B026 & 59061.533633287 & 20200731 & B051 & 59077.447490299 & 20200816-2 \\
    B002 & 58963.785093305 & 20200424 & B027 & 59061.533634057 & 20200731 & B052 & 59077.447490993 & 20200816-2 \\
    B003 & 58991.677837129 & 20200522 & B028 & 59061.534968368 & 20200731 & B053 & 59077.448090352 & 20200816-2 \\
    B004 & 58991.679623921 & 20200522 & B029 & 59061.536363277 & 20200731 & B054 & 59077.448491142 & 20200816-2 \\
    B005 & 58991.680362002 & 20200522 & B030 & 59061.538893001 & 20200731 & B055 & 59077.448491331 & 20200816-2 \\
    B006 & 58991.698122948 & 20200522 & B031 & 59067.465150308 & 20200806-1 & B056 & 59077.458653273 & 20200816-2 \\
    B007 & 58991.698124545 & 20200522 & B032 & 59067.484346541 & 20200806-2 & B057 & 59077.459056942 & 20200816-2 \\
    B008 & 58991.711183404 & 20200522 & B033 & 59067.484347120 & 20200806-2 & B058 & 59077.464848734 & 20200816-2 \\
    B009 & 58991.711364585 & 20200522 & B034 & 59067.500301102 & 20200806-2 & B059 & 59077.467268826 & 20200816-2 \\
    B010 & 58991.711703334 & 20200522 & B035 & 59067.507509019 & 20200806-2 & B060 & 59077.468457071 & 20200816-2 \\
    B011 & 58991.711703739 & 20200522 & B036 & 59067.532858711 & 20200806-2 & B061 & 59077.473602320 & 20200816-2 \\
    B012 & 58991.711704151 & 20200522 & B037 & 59069.493704768 & 20200808-2 & B062 & 59077.473884892 & 20200816-2 \\
    B013 & 58991.728936089 & 20200522 & B038 & 59069.498991812 & 20200808-2 & B063 & 59077.476002554 & 20200816-2 \\
    B014 & 58991.744126136 & 20200522 & B039 & 59069.512791794 & 20200808-2 & B064 & 59077.484007217 & 20200816-2 \\
    B015 & 58991.744126888 & 20200522 & B040 & 59071.470504066 & 20200810-2 & B065 & 59077.484007333 & 20200816-2 \\
    B016 & 59060.481445628 & 20200730-1 & B041 & 59071.470504297 & 20200810-2 & B066 & 59077.488582821 & 20200816-2 \\
    B017 & 59060.504831208 & 20200730-2 & B042 & 59071.489679757 & 20200810-2 & B067 & 59077.489969540 & 20200816-2 \\
    B018 & 59060.522934761 & 20200730-2 & B043 & 59073.495060445 & 20200812-2 & B068 & 59077.496363178 & 20200816-2 \\
    B019 & 59061.509817871 & 20200731 & B044 & 59073.513431187 & 20200812-2 & B069 & 59077.496514387 & 20200816-2 \\
    B020 & 59061.509818073 & 20200731 & B045 & 59073.513431997 & 20200812-2 & B070 & 59077.497517645 & 20200816-2 \\
    B021 & 59061.513340574 & 20200731 & B046 & 59075.452713325 & 20200814-2 & B071 & 59089.427874939 & 20200828 \\
    B022 & 59061.513341512 & 20200731 & B047 & 59075.453222841 & 20200814-2 & B072 & 59089.435877444 & 20200828 \\
    B023 & 59061.513342044 & 20200731 & B048 & 59075.470543041 & 20200814-2 & B073 & 59111.372277248 & 20200919-2 \\
    B024 & 59061.521404592 & 20200731 & B049 & 59075.482549483 & 20200814-2 & B074 & 59111.372277387 & 20200919-2 \\
    B025 & 59061.532697673 & 20200731 & B050 & 59075.494828132 & 20200814-2 & B075 & 59111.372729390 & 20200919-2 \\
    \enddata
    \tablecomments{\\
        $^a$ Topocentric MJD at 1.5 GHz.\\
        $^b$ In the observation details shown in Tab.~\ref{tab:obinfo} from FAST, some dates correspond to two separate observation sessions. In this table, `-1' and `-2' after the date indicate the first and second observation sessions on that day, respectively.
    }
\end{deluxetable}

\section{Dynamic Spectrum of Newly Detected Bursts from FRB 20190520B}

\begin{figure}[!htbp]
    \centering
    \includegraphics[width=0.9\textwidth]{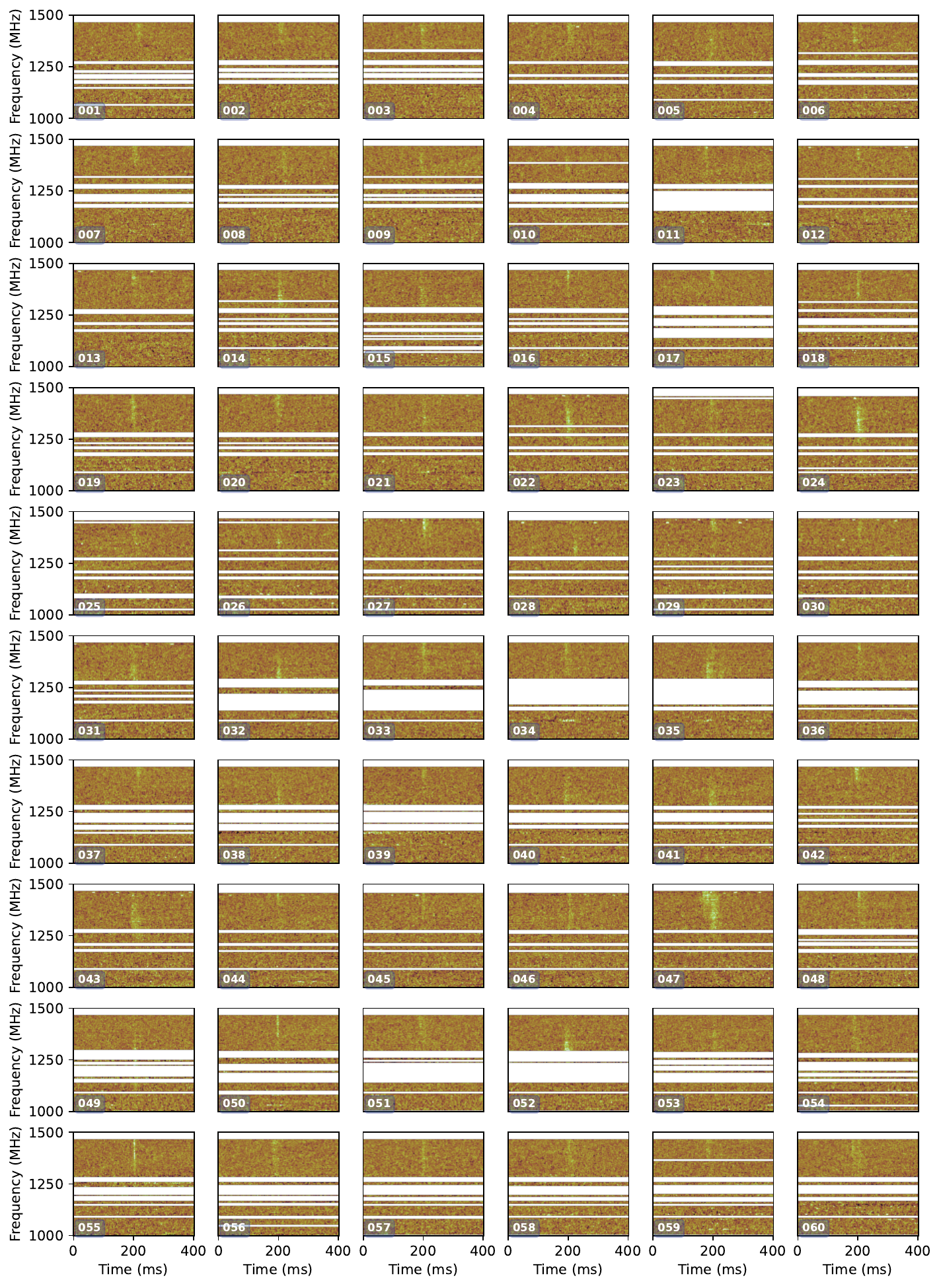}
    \caption{{\bf 183 newly detected bursts from FRB 20190520B.}}
    \label{fig:newburst}
\end{figure}

\addtocounter{figure}{-1}
\begin{figure}[!htbp]
    \centering
    \includegraphics[width=0.9\textwidth]{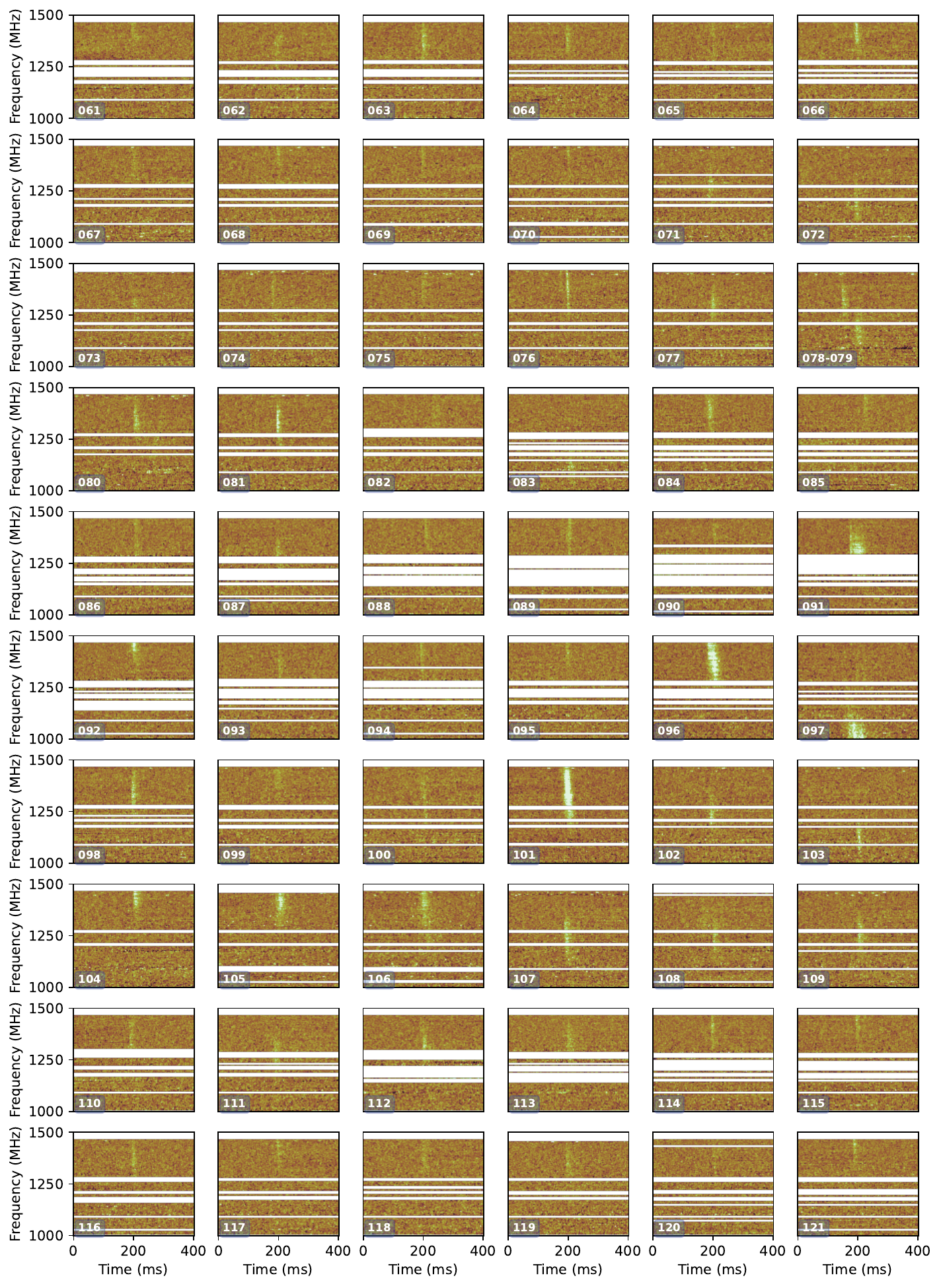}
    \caption{\textit{(continued)}}
\end{figure}

\addtocounter{figure}{-1}
\begin{figure}[!htbp]
    \centering
    \includegraphics[width=0.9\textwidth]{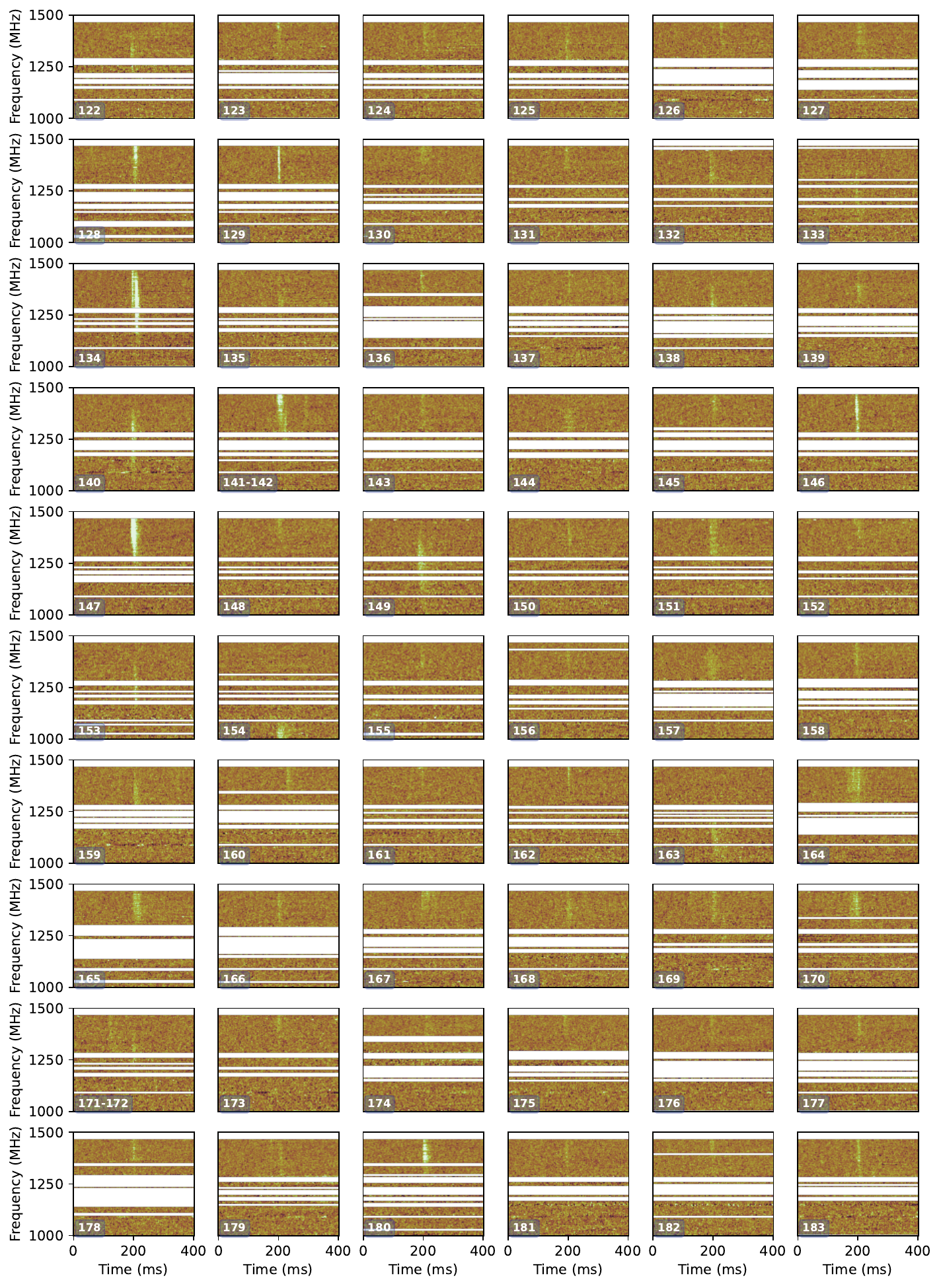}
    \caption{\textit{(continued)}}
\end{figure}

\bibliography{reference}{}
\bibliographystyle{aasjournal}

\end{document}